\documentclass[letterpaper]{jpconf}
\usepackage{graphicx}
\usepackage{amsmath}
\usepackage{amssymb}
\usepackage{citesort}
\RequirePackage{xspace}

\newcommand{\gev}{\ensuremath{{\mathrm{\,Ge\kern -0.1em V\!}}}\xspace}
\newcommand{\mev}{\ensuremath{{\mathrm{\,Me\kern -0.1em V\!}}}\xspace}
\newcommand{\gevc}{\ensuremath{{\mathrm{\,Ge\kern -0.1em V\!/}c}}\xspace}
\newcommand{\mevc}{\ensuremath{{\mathrm{\,Me\kern -0.1em V\!/}c}}\xspace}
\newcommand{\gevcc}{\ensuremath{{\mathrm{\,Ge\kern -0.1em V\!/}c^2}}\xspace}
\newcommand{\mevcc}{\ensuremath{{\mathrm{\,Me\kern -0.1em V\!/}c^2}}\xspace}
\newcommand{\tev}{\ensuremath{{\mathrm{\,Te\kern -0.1em V\!}}}\xspace}
\newcommand{\tevc}{\ensuremath{{\mathrm{\,Te\kern -0.1em V\!/}c}}\xspace}
\newcommand{\tevcc}{\ensuremath{{\mathrm{\,Te\kern -0.1em V\!/}c^2}}\xspace}

\newcommand{\dz}{\ensuremath{D{}^0}\xspace}

\newcommand{\dsp}{\ensuremath{D{}_{s}^{+}}\xspace}
\newcommand{\dssp}{\ensuremath{D{}_{(s)}^{+}}\xspace}
\newcommand{\dssm}{\ensuremath{D{}_{(s)}^{-}}\xspace}
\newcommand{\dpms}{\ensuremath{D^{*+}}\xspace}
\newcommand{\brll}[2]{\mathcal{B}\left(\dz\to#1^{+}#2^{-}\right)}

\begin{document}
\title{Search for $\dz \to \ell^+ \ell^-$ decays and for CP violation in $\dssp \to K_S^0 \pi^+$ and $\dssp\to K_S^0 K^+$ at BELLE}

\author{Marko Petri\v c (on behalf of the BELLE collaboration)}

\address{Jo\v zef Stefan Institute, Jamova cesta 39, 1000 Ljubljana, Slovenia}

\ead{marko.petric@ijs.si}

\begin{abstract}
We are reporting on a search for  flavour-changing neutral current decays $\dz \to \mu^+ \mu^-$ and $\dz \to e^+ e^-$, and for lepton-flavour violating decays $\dz \to e^\pm \mu^\mp$, the measurement of $\dssp \to K_S^0 \pi^+$ and $\dssp\to K_S^0 K^+$ branching fractions, and the search for CP violation in $\dssp \to K_S^0 \pi^+$ and $\dssp\to K_S^0 K^+$ decays. The analyses are based on 600 fb$^{-1}$ to 700 fb$^{-1}$ of data collected in $e^+e^-$ collisions at the centre-of-mass (CM) energy of the $\Upsilon(4S)$ resonance and 60~\mev below by the Belle detector at the KEKB collider.
\end{abstract}

\section{Search for $\dz \to \ell^+ \ell^-$ decays}

The flavour-changing neutral current (FCNC) decays $\dz\to~e^+e^-$ and 
$\dz\to\mu^+\mu^-$~\cite{cc} are highly suppressed in the standard model (SM) by the 
Glashow-Iliopoulos-Maiani mechanism~\cite{gim}. With the inclusion of long distance contributions the branching fractions can reach values of around $10^{-13}$~\cite{smfcnc}, these predictions are orders of magnitude below the current experimental sensitivity. The lepton-flavour violating (LFV) decays $\dz\to e^\pm\mu^\mp$ are forbidden in the SM.
In certain new physics scenarios, FCNC branching fractions can be enhanced by 
many orders of magnitude and FLV decays may become possible~\cite{newphys}.
For example, so far unobserved leptoquarks could enhance $\brll{\mu}{\mu}$ to $8\times 10^{-7}$~\cite{Dorsner:2009cu} .

Using $660\,\mbox{fb}^{-1}$ of data we searched for the decays $\dz\to\mu^+\mu^-$, 
$\dz\to e^+e^-$ and $\dz\to e^\pm\mu^\mp$. We use $\dz$ mesons from the decays $\dpms\to\dz\pi^{+}_{s}$ with a characteristic low momentum pion, since this considerably improves the purity of the reconstructed samples.  We normalise the sensitivity of our search to topologically similar $\dz\to\pi^{+}\pi^{-}$ decays; this cancels various systematic uncertainties. 
The signal efficiencies $\epsilon_{\ell\ell}$ and $\epsilon_{\pi\pi}$ are evaluated using signal Monte Carlo simulation.

In order to avoid biases, a blind analysis technique has 
been adopted. As the $\dz \to \ell^{+}\ell^{-}$ decays are not expected to be observed at the current sensitivity, we maximise the figure-of-merit, 
$\mathcal{F}=\epsilon_{\ell\ell}/N_{\rm UL}$, where $\epsilon_{\ell\ell}$ is the efficiency for 
detecting $D^0\to \ell^+\ell^-$ decays, and 
$N_{\rm UL}$ is the Poisson average of 
Feldman-Cousins 90\% confidence level upper limits on the number of 
observed signal events that would be obtained with the expected background 
and no signal~\cite{fc}.

The background events can be grouped into two categories: 
(1) a smooth combinatorial background, and (2) a 
peaking background from the misidentification of $\dz\to \pi^+\pi^-$ 
decays.
To estimate the number of combinatorial background events in the signal region,
the sideband region is used.  
The peaking background in the signal region due to misidentification 
of $\dz \to \pi^+\pi^-$ decays is estimated from the reconstructed $\dz \to \pi^+\pi^-$
decays found in data  and the misidentification  probability measured in data using $\dpms\to~\dz~\pi_{s}^+,\, \dz\to~K^{-}\pi^{+}$ decays, binned in particle momentum $p$ and cosine of polar angle. 
\begin{figure}[h]
\includegraphics[width=0.45 \columnwidth]{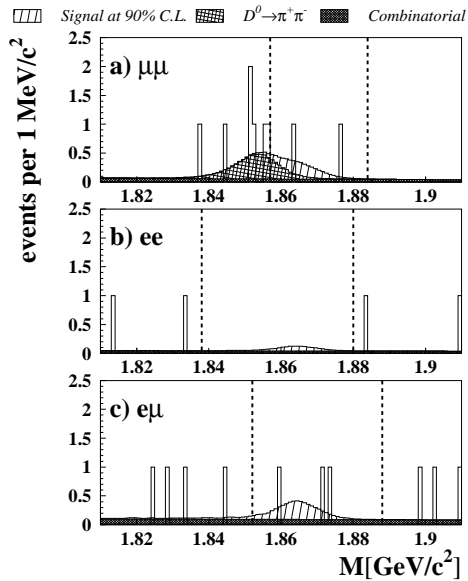}\hspace{0.0 \columnwidth}%
\begin{minipage}[b]{0.55 \columnwidth}\caption{\label{fig:Dtoll} The dilepton invariant mass distributions for a) $\dz \to \mu^+\mu^-$, b) $\dz \to e^+e^-$ and c) $\dz \to e^{\pm}\mu^{\mp}$. The dashed vertical lines indicate the optimised signal window. Superimposed on the data (open histograms) are the estimated distribution for combinatorial background (filled histogram), the misidentification of $\dz~\to~\pi^+\pi^-$ (cross-hatched histogram), and the signal if the branching fractions were equal to the 90\% confidence level upper limit (single hatched histogram).}
\end{minipage}
\end{figure}

The invariant mass distributions after applying 
the optimised event selection criteria are shown in Figure~\ref{fig:Dtoll}.
In the signal region we find
two candidates in the $\dz \to \mu^+\mu^-$, zero candidates in the $\dz \to e^+e^-$ and
three candidates in the $\dz \to e^{\pm}\mu^{\mp}$ decay mode; the yields are consistent
with the estimated background of $3.1\pm0.1$, $1.7\pm0.2$, and $2.6\pm0.2$ events respectively.
A binned maximum likelihood fit is used to determine the yield of $\dz \to \pi^{+}\pi^{-}$ candidates for the normalisation. 
Finally, the branching fraction upper limits (UL) are calculated using the program 
\texttt{pole.f}, which extends
the Feldman-Cousins method by the inclusion of systematic uncertainties~\cite{conrad}.
The upper limits on the branching fractions at the 90\% confidence level are found to be
$\mathcal{B}\left(\dz\to \mu^+\mu^-\right)<1.4\times 10^{-7}$, 
$\mathcal{B}\left(\dz\to e^+e^-\right)<7.9\times 10^{-8}$, and
$\mathcal{B}\left(\dz\to e^\pm\mu^\mp \right)<2.6\times 10^{-7}$~\cite{petric}.
Our results improve the current limits by a factor of 9 for $\dz\to \mu^+\mu^-$ decay, 
by a factor of 15 for $\dz\to e^+e^-$ decay and by a factor of 3 for
$\dz\to e^\pm\mu^\mp$ decay~\cite{PDG}. In 2008 the CDF collaboration reported a preliminary
result on the UL for the $\dz\to \mu^+\mu^-$ branching fraction~\cite{CDF}; 
our result is lower by a factor of 3 and strongly disfavours a leptoquark contribution~\cite{Dorsner:2009cu} as the explanation for the anomaly in the measured  $\dsp\to \mu^+ \nu$ width~\cite{Fajfer:2009qg}.

\section{Branching fraction measurement of $\dssp \to K_S^0 \pi^+$ and $\dssp\to K_S^0 K^+$ decays}
 Decays of charmed mesons play an important role 
in understanding the sources of SU(3) flavour symmetry breaking~\cite{ref:rosner}. 
For $D^+$ decays, the branching ratio 
$\mathcal{B}(D^+ \rightarrow \overline{K}^0 K^+)/\mathcal{B}(D^+ \rightarrow 
\overline{K}^0 \pi^+)$ deviates from the naive $\tan^2{\theta_C}$ 
expectation~\cite{PDG}, due to the destructive interference 
between colour-favoured and colour-suppressed amplitudes
in $D^+ \rightarrow \overline{K}^0 \pi^+$~\cite{ref:guberina}.
However, converting experimental measurements of $D$ decays that include
$K_S^0$ branching ratios to those involving $K^0$ or $\overline{K}^0$ is 
not straightforward due to the interference between the doubly Cabibbo-suppressed 
(DCS) and Cabibbo-favoured (CF) decay modes where
the interference phase is unknown~\cite{BIGI,ref:cleo_dksk}.

\begin{figure}[h]
\includegraphics[width=0.45\columnwidth]{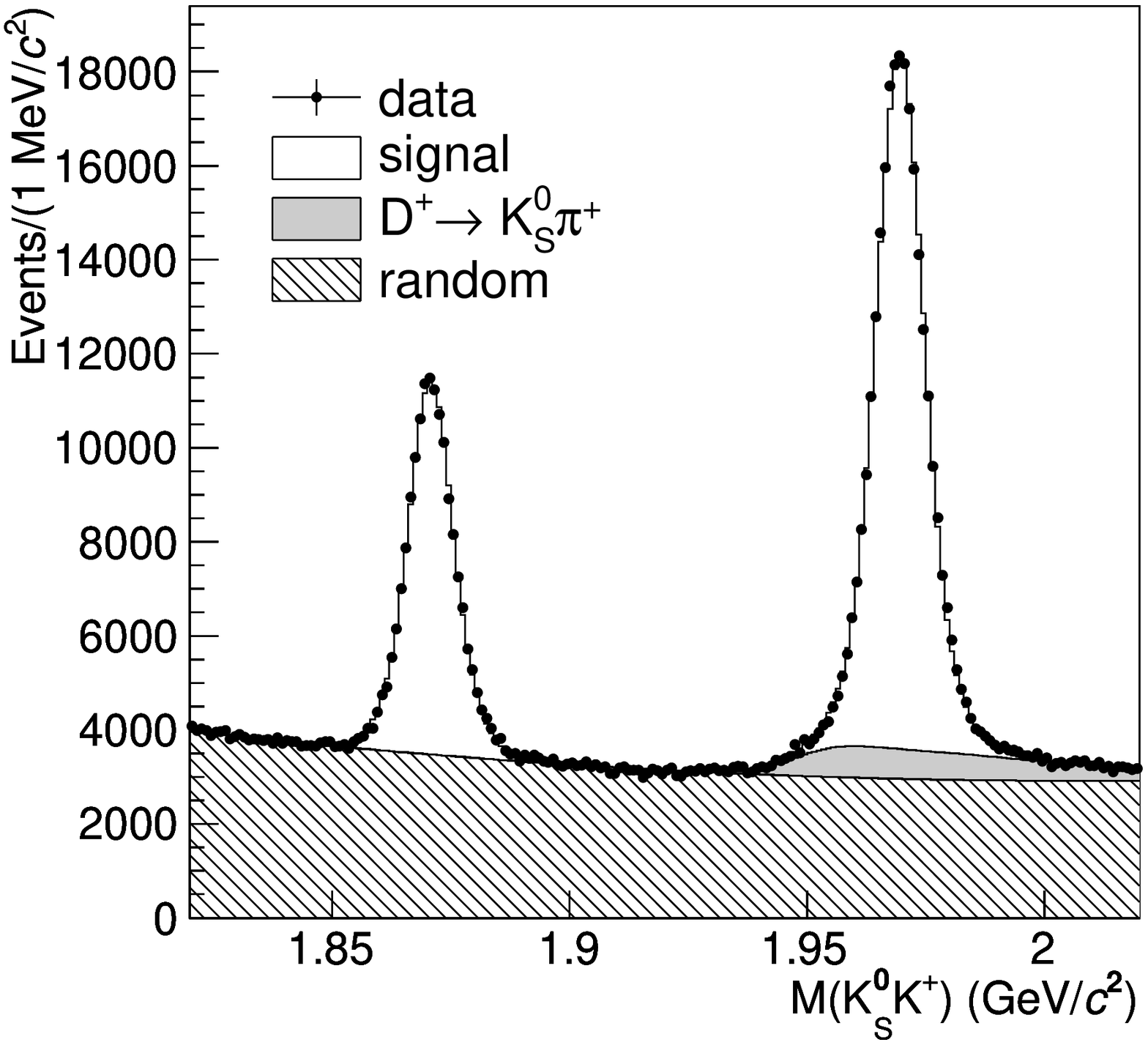}
\hspace{2pc}%
\includegraphics[width=0.45\columnwidth]{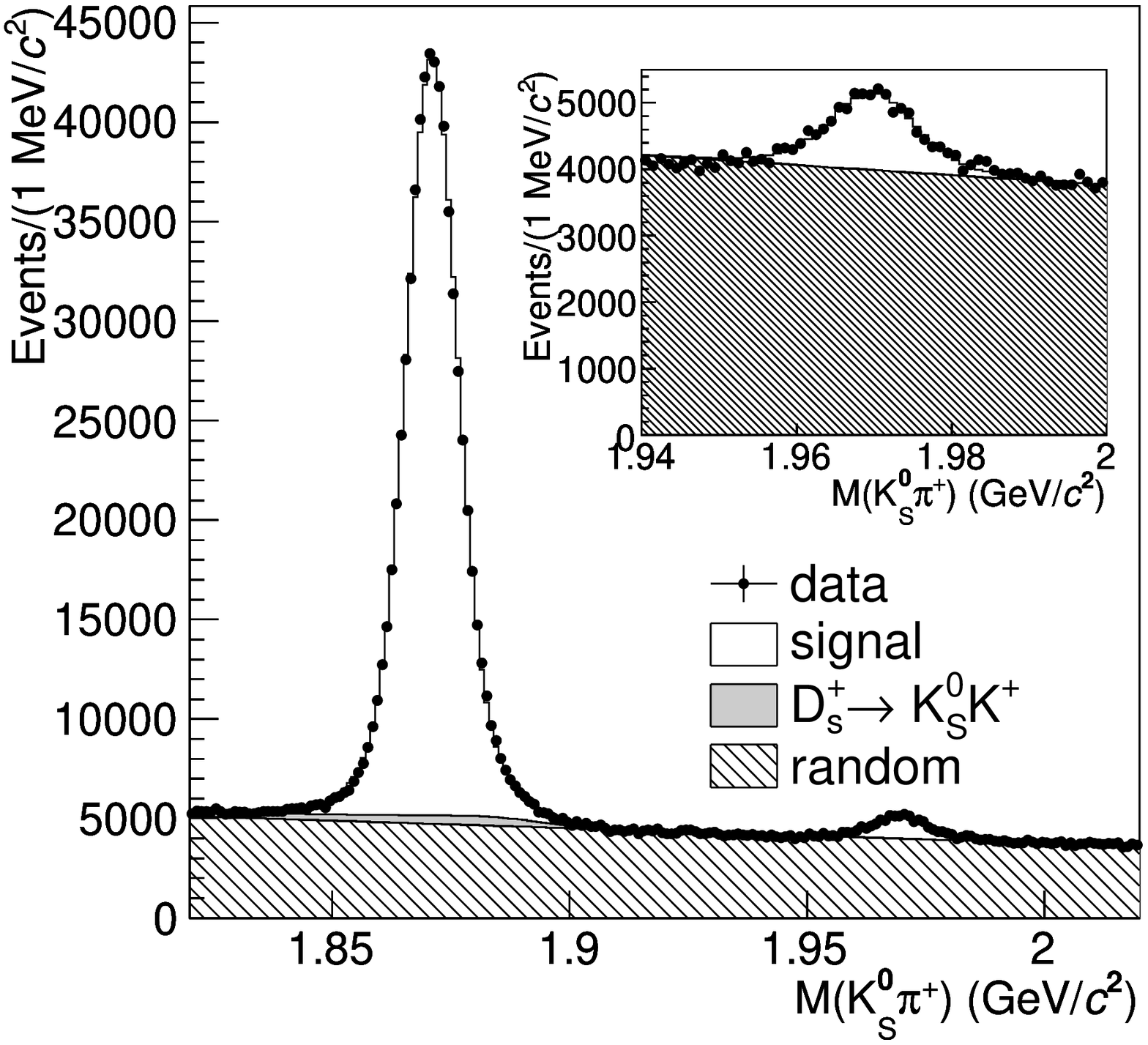}
\caption{\label{fig:data_ksk}Invariant mass distribution of 
selected
$K_S^0 K^+$ pairs {\sc left} and $K_S^0 \pi^+$ pairs {\sc right}. 
Points with error bars
show the data and histograms show the results of the fits. Signal, $D_{(s)}^+ \rightarrow K_S^0 h^+$ background ($h=K$ or $\pi$), and random
combinatorial background components are also shown.The inset is an enlarged view of the $D^+_s$ region.}
\end{figure}

Based on a data sample of 605 fb$^{-1}$ we measured the
$D^+ \rightarrow K_S^0 K^+$ and 
$D^+_s \rightarrow K_S^0 \pi^+$ branching ratios 
with respect to the corresponding Cabibbo-favoured modes. The invariant mass distributions of the selected events are shown in Figure~\ref{fig:data_ksk}. The results are
$\mathcal{B}(D^+ \rightarrow K_S^0 K^+)/\mathcal{B}(D^+ \rightarrow K_S^0 \pi^+)$ = 0.1899$\pm$0.0011$\pm$0.0022 
and 
$\mathcal{B}(D^+_s \rightarrow K_S^0 \pi^+)/\mathcal{B}(D^+_s \rightarrow K_S^0 K^+)$ = 0.0803$\pm$0.0024$\pm$ 0.0019, 
where the first uncertainties are statistical and the second are systematic~\cite{brkorat}.
Using the world average values of CF decay rates~\cite{PDG}, we obtain 
the branching fractions 
$\mathcal{B}(D^+ \rightarrow K_S^0 K^+)$ = (2.75$\pm$0.08)$\times$10$^{-3}$ 
and 
$\mathcal{B}(D^+_s \rightarrow K_S^0 \pi^+)$ = (1.20$\pm$0.09)$\times$10$^{-3}$ 
where the uncertainties are the sum in quadrature of statistical and 
systematic errors.
These are consistent with the present world averages~\cite{PDG} 
and are the most precise measurements up to now. The ratio 
$\mathcal{B}(D^+ \rightarrow K_S^0 K^+)$/$\mathcal{B}(D^+_s \rightarrow K_S^0 \pi^+)$ = 2.29$\pm$0.18  
may be due to SU(3) flavour breaking and/or different 
final-state interactions in $D^+$ and $D^+_s$ decays.

\section{Search for CP violation in $\dssp \to K_S^0 \pi^+$ and $\dssp\to K_S^0 K^+$ decays}
Another important aspect of such decays is the violation of the combined Charge-conjugation and Parity symmetries ($CP$).
In the SM, the charmed particle processes for which a significant non-vanishing
$CP$ violation is expected are singly Cabibbo-suppressed (SCS) decays in which
there is both interference between two different decay amplitudes and a strong
phase shift from final state interactions. In the SM, $CP$ violation in SCS
charmed meson decays is predicted to occur at the level of $\mathcal{O}(0.1)$\%
or lower~\cite{SMCP}. 

Based on a data sample of 673 fb$^{-1}$ we determine the $CP$ violating asymmetry $A_{CP}$ by measuring the signal yield asymmetry
$
  A_{\rm rec}=(
  N_{\rm rec}-\overline N_{\rm rec})/(
  N_{\rm rec}+\overline N_{\rm rec})
$
where $N_{\rm rec}$($\overline N_{\rm rec}$) is the number of reconstructed decays of $\dssp(\dssm)$. The measured asymmetry in this equation includes two
contributions other than $A_{CP}$. One is the forward-backward asymmetry
($A_{FB}$) due to $\gamma^{*}-Z^0$ interference in $e^+e^-\rightarrow c\bar{c}$
and the other is a detection efficiency asymmetry between positively and
negatively charged tracks
$A^h_{\epsilon}=(\epsilon^+ - \epsilon^-)/(\epsilon^+ +  \epsilon^-)$, where $\epsilon^+(\epsilon^-)$ 
is the efficiency for $K^+(K^-)$ or $\pi^+(\pi^-)$ meson and $h$ denotes $K$ or $\pi$. Since $K^0_S$ mesons are
reconstructed from a $\pi^+\pi^-$ pair, there is no detection asymmetry other
than $A^{h}_{\epsilon}$. The signal yield asymmetry can therefore be expressed as
$
  A_{\rm rec}=A_{CP}+A_{FB}+A^{h}_{\epsilon}.
$

To correct for the asymmetries other than $A_{CP}$, we use reconstructed
samples of $D^+_s\rightarrow\phi\pi^+$ and $D^0\rightarrow K^-\pi^+$ decays and
assume that $A_{CP}$ in CF decays is negligibly small compared to the current
experimental sensitivity and that $A_{FB}$ is the same for all charmed
mesons. We reconstruct $\phi$ mesons via their $\phi\to K^+K^-$ decays. 
The measured asymmetry for $D^+_s\rightarrow\phi\pi^+$ is the sum of
$A_{FB}$ and $A^{\pi}_{\epsilon}$. Hence one can extract the $A_{CP}$
value for the $K^0_S\pi^+$ final state by subtracting the measured asymmetry for
$D^+_s\rightarrow\phi\pi^+$ from that for $D^+_{(s)}\rightarrow K^0_S\pi^+$. 

\begin{figure}[h]
\includegraphics[width=0.5 \columnwidth]{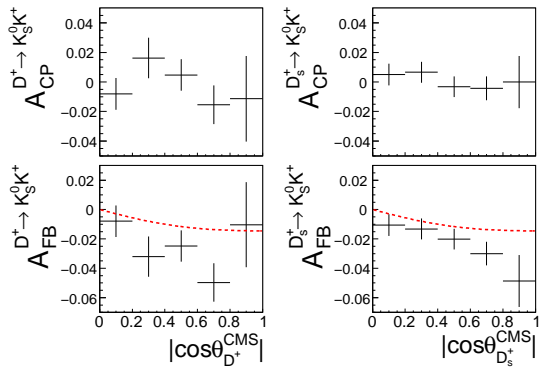}\hspace{0.000 \columnwidth}%
\begin{minipage}[b]{0.5 \columnwidth}\caption{\label{FIG:ACPKSK}Measured $A_{CP}$ and $A_{FB}$ values for $D^+_{(s)}\rightarrow K^0_S K^+$ as
  a function of $|\cos\theta^{\rm CMS}_{D^+_{(s)}}|$. The dashed curves show
  the leading-order prediction for $A^{c\bar{c}}_{FB}$.}
\end{minipage}
\end{figure}

The method for the measurement of $A_{CP}$ in the $K^0_S K^+$ final states is
different from that for the $K^0_S\pi^+$ final states. The $A_{FB}$
and $A^{\pi}_{\epsilon}$ components in $A_{\rm rec}$ are directly obtained from the
$D^+_s\rightarrow\phi\pi^+$ sample, but there is no corresponding large
statistics decay mode that can be used to directly measure the
$A_{FB}$ and $A^{K}_{\epsilon}$ components in
$A_{\rm rec}$. Thus, to correct the
reconstructed asymmetry in the $K^0_S K^+$ final states, we use samples of
$D^0\rightarrow K^-\pi^+$ as well as $D^+_s\rightarrow\phi\pi^+$ decays.
The value $A_{\rm rec}-A^{K}_{\epsilon}$ includes not only an $A_{CP}$ component but also an $A_{FB}$ component. Since $A_{CP}$ is independent of all kinematic
    variables, while $A_{FB}$ is an odd function of $\cos\theta^{\rm
    CMS}_{D^+_{(s)}}$, we can deduce both by addition/subtraction in bins of $\cos\theta$.
Figure~\ref{FIG:ACPKSK} shows the results.

No evidence for $CP$
violation has been observed~\cite{brkocp}. Our results are
$A^{D^+\rightarrow K^0_S\pi^+}_{CP}=(-0.71\pm0.19\pm0.20)\%$, 
$A^{D^+_s\rightarrow K^0_S\pi^+}_{CP}=(+5.45\pm2.50\pm0.33)\%$, 
$A^{D^+\rightarrow K^0_S K^+}_{CP}=(-0.16\pm0.58 \pm0.25)\%$, and
$A^{D^+_s\rightarrow K^0_S K^+}_{CP}=(+0.12\pm0.36\pm0.22)\%$. They are consistent with the SM predictions and provide the most stringent constraints up to now on models beyond the SM~\cite{BIGI}.

\section*{References}
\bibliography{llwi_petric}

\begin{thebibliography}{10}
\bibitem{cc}
Throughout this paper charge-conjugate modes are included
\end{thebibliography}
\end{document}